\documentclass[12pt]{article}
\usepackage[dvips]{graphics}
\usepackage{graphicx}
\title{Superluminal velocities and nonlocality in relativistic mechanics with scalar potential}
\author{Hrvoje Nikoli\'c \\
Theoretical Physics Division, Rudjer Bo\v{s}kovi\'{c} Institute, \\
P.O.B. 180, HR-10002 Zagreb, Croatia \\
{\normalsize e-mail: hrvoje@thphys.irb.hr} \\
\makebox[1in]{} \\
}
\date{\today}
\begin{document}
\maketitle

\begin{abstract}
Even though the usual form of relativistic mechanics does not allow
superluminal particle velocities and nonlocal interactions,
these features are not forbidden by relativity itself. To understand this
on a deeper level, we study a generalized form of relativistic mechanics 
in which the particle is influenced not only by the usual 
tensor (gravitational) and vector (electromagnetic) potentials, but also
by the scalar potential. The scalar potential promotes the mass
squared $M^2$ to a dynamical quantity. Negative values of $M^2$,
which lead to superluminal velocities, are allowed. 
The generalization to the many-particle case allows a nonlocal scalar
potential, which makes nonlocal interactions compatible with
relativity. 
Particle trajectories are parameterized by a scalar parameter analogous
to the Newton absolute time.
An example in which all these general features
are explicitly realized is provided by relativistic Bohmian mechanics.
\end{abstract}
 
\vspace*{0.5cm}

\section{Introduction}

In physics literature, one can frequently find a statement that the theory of
relativity implies that particles cannot move faster than light. Yet, this is not really true. 
The theory of relativity allows tachyons \cite{tach1,tach2,tach3} -- 
hypothetical particles with negative squared mass -- that move only 
with {\em superluminal} velocities, i.e., velocities faster than light. 
Similarly, it is frequently stated that relativity allows only local interactions, 
which is incorrect too.

In this paper we systematically explore the possibility
of superluminal velocities and nonlocal interactions 
in relativistic mechanics of classical particles from a theoretically deeper
and more general 
point of view. We develop a generalized form of relativistic mechanics 
in which the particle is influenced not only by the usual 
tensor (gravitational) and vector (electromagnetic) potentials, but also
by the scalar potential. 
Inspired by the formalism popular in the string-theory community 
\cite{string1,string2,string3,string4}, we describe dynamics by the action
invariant with respect to general reparameterizations of the
particle trajectory.
It turns out that
the scalar potential promotes the mass squared to a dynamical quantity
which may change its sign. It also turns out that the scalar potential
may describe nonlocal interactions between particles in a relativistic-covariant
manner. The manifestly relativistic-covariant formalism turns out
to be analogous to the formalism of nonrelativistic Newtonian mechanics. 
In particular, the scalar
parameter that parameterizes relativistic particle trajectories turns out to be
analogous to the Newton absolute time.

The possibility of superluminal velocities and nonlocal interactions within 
classical relativistic mechanics is interesting in its own right.
Yet, this possibility is even more interesting from the point of view
of the Bohmian formulation of quantum mechanics 
\cite{BMbook1,BMbook2,BMbook3,BMbook4,BMbook5,BMbook6}. 
In this formulation, particles have classical-like trajectories that
interact with each other in a nonlocal manner. Relativistic-covariant
equations of motion for the trajectories have been studied in
\cite{durr96,nik3,nik4,hern}. The corresponding relativistic-covariant 
probabilistic interpretation and its compatibility with 
the predictions of standard
quantum theory has been discussed in  
\cite{nikbosfer,nikijqi,nikqft,niktorino}.
In this paper we show that the relativistic-covariant
equations for Bohmian trajectories of spin-0 particles
can be viewed as a special case of the
general theory of classical relativistic mechanics.

The paper is organized as follows. After a brief review of some well-known aspects 
of Newtonian mechanics in Sec.~\ref{SEC2}, we present the basics
of relativistic kinematics in Sec.~\ref{SEC3}. Then, in the central section of this paper, 
Sec.~\ref{nikolic:Dynamics}, we develop the general formalism of relativistic
dynamics. In Sec.~\ref{SEC5} we study the case of relativistic Bohmian mechanics, as an
example of the general formalism developed in Sec.~\ref{nikolic:Dynamics}.
Finally, the conclusions are drawn in Sec.~\ref{SEC6}.

\section{Preliminaries: Aspects of Newtonian mechanics}
\label{SEC2}

Relativistic mechanics, which will be studied in the subsequent sections,
has many formal and conceptual similarities with nonrelativistic
Newtonian mechanics. Thus, to make relativistic mechanics easier to comprehend, 
in this section we remind the reader
of some well-known aspects of Newtonian mechanics 
(see, e.g., \cite{goldstein})
which will turn out to have 
a direct analogy in relativistic mechanics.

Consider $n$ particles with trajectories ${\bf X}_a(t)$,
$a=1,\ldots,n$,  that interact with each other by Newton gravitational force.
For simplicity, we assume that all particles have the same mass $m$.
Then the interaction is described by the nonlocal Newton potential
\begin{equation}\label{poten}
 U({\bf x}_1,\ldots,{\bf x}_n)=\frac{1}{2}\sum_a \sum_{a'\neq a}
\frac{-m^2G_{\rm N}}{|{\bf x}_a-{\bf x}_{a'}|} ,
\end{equation}
where $G_{\rm N}$ is the Newton constant.
Particle trajectories satisfy the Newton equation
\begin{equation}\label{Newton}
 m\frac{d^2X^i_a(t)}{dt^2}=-\partial^i_a U({\bf X}_1(t),\ldots,{\bf X}_n(t)) ,
\end{equation}
where $\partial^i_a \equiv \partial/\partial x^i_a$ and $i=1,2,3$ labels
the 3 Cartesian space coordinates of ${\bf x}\equiv (x^1,x^2,x^3)$.

The Hamiltonian for the system above is
\begin{equation}\label{Ham}
 H({\bf P}_1,\ldots,{\bf P}_n,{\bf X}_1,\ldots,{\bf X}_n)
=\sum_a \frac{{\bf P}_a^2}{2m} + U({\bf X}_1,\ldots,{\bf X}_n),
\end{equation}
where ${\bf P}_a$ are the particle momenta
\begin{equation}\label{momenta}
 {\bf P}_a=m\dot{{\bf X}}_a ,
\end{equation}
and $\dot{{\bf X}}_a \equiv d {\bf X}_a(t)/dt$ are velocities.
The dynamics of this system can be described by the Hamilton-Jacobi
formalism. In this formalism, one first needs to find the solution
$S({\bf x}_1,\ldots,{\bf x}_n,t)$ of the Hamilton-Jacobi equation
\begin{equation}\label{HamJac}
 H(\nabla_1S,\ldots,\nabla_nS,{\bf x}_1,\ldots,{\bf x}_n) = -\frac{\partial S}{\partial t} ,
\end{equation}
where $\nabla_aS\equiv(\partial^1_aS,\partial^2_aS,\partial^3_aS)$. 
Then the particle trajectories are given by (\ref{momenta}), or more precisely by
\begin{equation}\label{traj}
 m\frac{dX^i_a(t)}{dt}=\partial^i_aS({\bf X}_1(t),\ldots,{\bf X}_n(t)) .
\end{equation}

The Hamiltonian represents the total energy of the system. This energy is conserved.
But where energy comes from in the first place? Newtonian mechanics does not
provide an answer to this question. Yet, one possibility is that there is no energy at all,
i.e., that total energy is zero
\begin{equation}\label{Hamconstr}
 H=0 .
\end{equation}
Since the potential energy (\ref{poten}) is negative, the Hamiltonian constraint
(\ref{Hamconstr}) is not trivial. Instead, particles may exhibit a complicated motion
even when the total energy vanishes. (In fact, in 
the general-relativistic theory of gravity this is always
the case. More precisely, 
in general relativity the sum of gravitational energy density and matter energy density is 
zero everywhere \cite{weinberg}.)
For a later comparison with relativistic mechanics, we also note that the
Hamiltonian constraint (\ref{Hamconstr}), combined with (\ref{Ham}) and
(\ref{momenta}), can be written as
\begin{equation}\label{dt2}
dt^2=-\frac{m}{2U({\bf X}_1,\ldots,{\bf X}_n)} \sum_a d{\bf X}_a^2 . 
 \end{equation}

Finally, let us give a few conceptual remarks on the Newton time $t$. 
It is an absolute, observer-independent time. Also,
it is an external parameter on which the Hamiltonian (\ref{Ham}) does 
not depend explicitly. In fact, the only role of $t$ in the equations above is to
parameterize the particle trajectories in space. In this sense, $t$ is only an
auxiliary parameter; it is not directly measurable. Yet, a ``clock'' can measure 
$t$ indirectly. Namely, a ``clock'' is a physical process periodic in time.
Hence, if at least one of the $3n$ functions
$X^i_a(t)$ is periodic in $t$ (e.g., a pendulum
driven by the gravitational force), 
then the number of periods can be thought of as a measure of elapsed time $t$.
This is how time is measured in practice.

\section{Relativistic kinematics}
\label{SEC3}

In relativistic physics, time is treated on an equal footing with space.
Instead of dealing with a separate notions of space and time, relativity is formulated
on a 4-dimensional spacetime with coordinates $x^{\mu}$,
$\mu=0,1,2,3$, and the Minkowski metric
$\eta_{\mu\nu}$, where $\eta_{00}=1$, $\eta_{ij}=-\delta_{ij}$, $\eta_{0i}=0$,
for $i=1,2,3$. We work in units in which the velocity of light is $c\equiv 1$.
The relation with nonrelativistic (Newtonian) physics is established through 
the notation $x^{\mu}=(t,{\bf x})$, where
$t=x^0$ is the time coordinate and ${\bf x}=(x^1,x^2,x^3)$ represents the
space coordinates.

The physical objects that we study are particles living in spacetime. By a particle 
we mean a material point in space. 
More precisely, since the concept of space is not a well-defined entity 
in relativity, a better definition of a particle is a curve in spacetime. Thus, 
the particle is a 1-dimensional object living in the 4-dimensional spacetime.
 
The simplest way to specify a curve is through a set of 4 equations
\begin{equation}\label{nikolic:book1}
x^{\mu}=X^{\mu}(s) ,
\end{equation}
where $s$ is an auxiliary real parameter and $X^{\mu}(s)$ are some 
specified functions of $s$. Each $s$ defines one point on the curve and the set
of all values of $s$ defines the whole curve. In this sense, the curve can be identified
with the functions $X^{\mu}(s)$.

The parameter $s$ is a scalar with respect to Lorentz transformations or any other transformations
of the spacetime coordinates $x^{\mu}$. In this sense, the parametric definition
of the curve (\ref{nikolic:book1}) is covariant. However, non-covariant definitions are also possible.
For example, if the function $X^{0}(s)$ can be inverted, then the inverse $s(X^0)$
can be plugged into the space components  $X^{i}(s(X^0))\equiv \tilde{X}^{i}(X^0)$.
This leads to the usual nonrelativistic view of the particle as an object with the trajectory
$x^i=\tilde{X}^{i}(X^0)$, where $X^0$ is time.

{\it A priori}, the auxiliary parameter $s$ does not have any physical interpretation.
It is merely a mathematical parameter that cannot be measured. In fact, a transformation 
of the form
\begin{equation}\label{nikolic:book2}
 s\rightarrow s'=f(s) 
\end{equation}
does not change the curve in spacetime. (The only restriction on the function
$f(s)$ is that $df(s)/ds >0$.) This means that the functions $X^{\mu}(s)$
and $\tilde{X}^{\mu}(s)\equiv X^{\mu}(f(s))$ represent the same curve.

Since the curve is a 1-dimensional manifold, the parameter $s$ can be viewed as a
coordinate on that manifold. The transformation (\ref{nikolic:book2}) is a 
coodinate transformation on that manifold. One can also define the metric
tensor $h(s)$ on that manifold, such that $h(s)ds^2$ is the (squared) infinitesimal
length of the curve. Since the manifold is 1-dimensional, the metric tensor
$h$ has only 1 component. It is important to stress that this is an
intrinsic definition of the length of the curve that may be defined completely
independently on the spacetime metric $\eta_{\mu\nu}$.
This intrinsic length is not measurable so one can freely choose the metric $h(s)$.
However, once $h(s)$ is chosen, the metric in any other coordinate $s'$ is defined
through
\begin{equation}\label{nikolic:book3}
 h(s)ds^2 = h'(s')ds'^2 .
\end{equation}

We say that the curve at a point $s$ is timelike if the spacetime vector tangent to the curve at this
point is timelike. Spacelike and lightlike parts of the curve are defined analogously.
Thus, the part of the curve is timelike if $\dot{X}^{\mu} \dot{X}_{\mu}>0$,
spacelike if $\dot{X}^{\mu} \dot{X}_{\mu}<0$, and lightlike if $\dot{X}^{\mu} \dot{X}_{\mu}=0$, where $\dot{X}^{\mu}=dX^{\mu}(s)/ds$. (In this paper 
$A^{\mu} B_{\mu}\equiv \eta_{\mu\nu}A^{\mu} B^{\nu}$ and the summation
over repeated vector indices $\mu$, $\nu$ is understood.)
A timelike trajectory describes a particle that moves slower than light, a lightlike trajectory
describes a particle that moves with the velocity of light, and a spacelike
trajectory describes a particle that moves faster than light.
Contrary to what one might expect, we see that relativistic kinematics allows
particles to move even faster than light. As we shall see in the next section,
it is relativistic dynamics that may (or may not!) forbid motions faster than light,
depending on details of the dynamics.

For a timelike trajectory, there exists one special choice of the parameter $s$.
Namely, one can choose it to be equal to the proper time $\tau$ defined by
\begin{equation}\label{nikolic:book4}
 d\tau^2=dX^{\mu}dX_{\mu} .
\end{equation}
For such a choice, we see that
\begin{equation}\label{nikolic:book5}
\dot{X}^{\mu} \dot{X}_{\mu}=1 .
\end{equation}
In this case it is convenient to choose the metric on the trajectory such that
$h(\tau)=1$, so that the intrinsic length of the curve coincides with the
proper time, which, by definition, is equal to the extrinsic length defined by the
spacetime metric $\eta_{\mu\nu}$. Yet, such a choice is by no means
necessary.

Finally, let us briefly generalize the results above to the case of many particles.
If there are $n$ particles, then they are described by $n$ trajectories
$X_a^{\mu}(s_a)$, $a=1,\ldots,n$. Note that each trajectory is parameterized by its own
parameter $s_a$. However, since the parameterization of each curve is arbitrary,
one may parameterize all trajectories by the same parameter $s$, so that
the trajectories are described by the functions $X_a^{\mu}(s)$. 
In fact, the functions $X_a^{\mu}(s)$, which describe $n$ curves in the
4-dimensional spacetime, can also be viewed as {\em one} curve on a $4n$-dimensional
manifold with coordinates $x_a^{\mu}$. 

\section{Relativistic dynamics}
\label{nikolic:Dynamics}

\subsection{Action and equations of motion}

Dynamics of a relativistic particle is described by an action of the form
\begin{equation}\label{nikolic:book6}
 A=\int ds \, L(X(s),\dot{X}(s),s) ,
\end{equation}
where $X\equiv \{ X^{\mu} \}$, $\dot{X}\equiv \{ \dot{X}^{\mu} \}$.
We require that the Lagrangian $L$ should be a scalar with respect to spacetime coordinate
transformations. This means that all spacetime indices $\mu$ must be contracted.
We also require that the action should be invariant with respect to
reparameterizations of the form of (\ref{nikolic:book2}). From (\ref{nikolic:book3}), we see that this 
implies that any $ds$ should by multiplied by $\sqrt{h(s)}$, because such a product
is invariant with respect to (\ref{nikolic:book2}). To restrict the dependence on $s$
as much as possible, we assume that there is no other explicit dependence on $s$
except through the dependence on $h(s)$.
To further restrict the possible forms
of the action, we require that $L$ should be at most quadratic in the velocities 
$\dot{X}^{\mu}(s)$. With these requirements, the most general action can be written
in the form
\begin{equation}\label{nikolic:book7}
 A=-\int ds \, \sqrt{h(s)} \left[ \frac{1}{2h(s)}  
\frac{dX^\mu}{ds} \frac{dX^\nu}{ds} C_{\mu\nu}(X)
+\frac{1}{\sqrt{h(s)}} \frac{dX^\mu}{ds} C_{\mu}(X) + C(X) \right] .
\end{equation}
The functions $C(X)$, $C_{\mu}(X)$, and $C_{\mu\nu}(X)$ are referred to as
scalar potential, vector potential, and tensor potential, respectively.

What is the dynamical role of the function $h(s)$? Requiring that $h(s)$ is a
dynamical variable, the dynamical equation of motion $\delta A/\delta h(s)=0$
leads to
\begin{equation}\label{nikolic:book8}
 h^{-1}C_{\mu\nu}(X)\dot{X}^{\mu}\dot{X}^{\nu}=2C(X) .
\end{equation}
Viewed as an equation for $h$, it can be trivially solved as 
$h=C_{\mu\nu}\dot{X}^{\mu}\dot{X}^{\nu}/2C$. However,
since $h$ is not a physical quantity, this solution does not bring an important
physical information. Nevertheless, Eq.~(\ref{nikolic:book8}) does play an important
physical role, as we shall see soon.

Eq.~(\ref{nikolic:book8}) determines $h(s)$ only when the coordinate $s$ is chosen.
Thus, $h(s)$ can still be changed by changing the coordinate. In particular,
from (\ref{nikolic:book3}) we see that the coordinate transformation of the form
$s'(s)={\rm const}\int ds \, \sqrt{h(s)}$ makes $h'(s')$ a constant.
Thus, omitting the prime, we can fix 
\begin{equation}\label{fix}
\sqrt{h(s)}=m^{-1},
\end{equation} 
where $m$ is a positive constant.
For convenience, we choose $s$ to have the dimension of time and $C_{\mu\nu}$ 
to be dimensionless. Then the action (\ref{nikolic:book7}) implies that $m$ has the dimension
of mass (recall that we work in units $c=1$). Hence, we can rewrite (\ref{nikolic:book7})
as
\begin{equation}\label{nikolic:book9}
 A=-\int ds \left[ \frac{m}{2}  C_{\mu\nu}(X)
\dot{X}^{\mu} \dot{X}^{\nu} 
+C_{\mu}(X) \dot{X}^{\mu} + \frac{C(X)}{m} \right] .
\end{equation}
Now $m$ is no longer a dynamical quantity, but Eq.~(\ref{nikolic:book8}) rewritten as
\begin{equation}\label{nikolic:book10}
 C_{\mu\nu}(X)\dot{X}^{\mu}\dot{X}^{\nu}=\frac{2C(X)}{m^2} 
\end{equation}
should be added to (\ref{nikolic:book9}) as an additional constraint.

Now we are ready to study the physical role of the potentials $C$, $C_{\mu}$ and $C_{\mu\nu}$.
By writing $C_{\mu}(x)\equiv e A_{\mu}(x)$, one recognizes that the second term in
(\ref{nikolic:book9}) looks just like the action for the particle with the charge $e$ 
moving under the influence of the external electromagnetic potential $A_{\mu}(x)$
(see, e.g., \cite{jackson}). 
Similarly, by writing $C_{\mu\nu}(x)\equiv g_{\mu\nu}(x)$, one recognizes that the first term in
(\ref{nikolic:book9}) looks just like the action for the particle moving in a gravitational
background described by the curved metric tensor $g_{\mu\nu}(x)$ (see, e.g., \cite{weinberg}).
Since the physical properties of electromagnetic and gravitational forces are 
well known, we shall not study them in further discussions. Instead, from now on
we assume $C_{\mu}(x)=0$, $C_{\mu\nu}(x)=\eta_{\mu\nu}$.
Therefore, introducing the notation $U(X)\equiv C(X)/m$,
Eqs.~(\ref{nikolic:book9}) and (\ref{nikolic:book10}) reduce to
\begin{equation}\label{nikolic:book11}
 A=-\int ds \left[ \frac{m}{2} \dot{X}^{\mu} \dot{X}_{\mu} + U(X) \right]  ,
\end{equation}
\begin{equation}\label{nikolic:book12}
 \dot{X}^{\mu}\dot{X}_{\mu}=\frac{2U(X)}{m} .
\end{equation}
We see that the scalar potential $U(X)$ has the dimension of energy.
The dynamical equation of motion for $X^{\mu}(s)$ is 
$\delta A/\delta X^{\mu}(s)=0$. Applying this to (\ref{nikolic:book11}), one obtains
a relativistic Newton equation
\begin{equation}\label{nikolic:book13}
 m\frac{d^2X^{\mu}(s)}{ds^2}=\partial^{\mu}U(X(s)) ,
\end{equation}
where $\partial^{\mu}\equiv \eta^{\mu\nu}\partial/\partial X^{\nu}$.
The constraint (\ref{nikolic:book12}) is compatible with (\ref{nikolic:book13}).
Indeed, by applying the derivative $d/ds$ on (\ref{nikolic:book12}), one obtains
\begin{equation}
 [m\ddot{X}^{\mu}-\partial^{\mu}U(X)] \dot{X}_{\mu}=0 ,
\end{equation}
which is consistent because the expression in the square bracket trivially vanishes
when (\ref{nikolic:book13}) is satisfied.

The constraint (\ref{nikolic:book12}) implies
that the sign of $\dot{X}^{\mu}\dot{X}_{\mu}$ is equal to the sign of
$U$. Thus, we see that the particle moves slower than light if $U>0$,
with the velocity of light if $U=0$, and faster than light if $U<0$. 
Since $U(X)$ may change sign as $X$ varies, we see that the particle may, e.g., 
start motion with a velocity slower than light and {\em accelerate to a velocity
faster than light}.

At first sight, one may think that acceleration to velocities faster than light
is in contradiction with the well-known ``fact'' that the principle of relativity
does not allow particles to accelerate to velocities faster than light. 
However, there is no contradiction because this well-known ``fact'' is valid
only if some additional assumptions are fulfilled. In particular, if all forces
on particles are either of the electromagnetic type (vector potential) 
or of the gravitational type (tensor potential), then acceleration to velocities
faster than light is forbidden. Indeed, as far as we know, 
all relativistic classical forces on particles that exist in nature are of those two types.
Nevertheless, the principle of relativity allows also relativistic forces based
on the scalar potential, which, as we have seen, does allow acceleration to velocities
faster than light. Such classical forces have not yet been found in nature,
but it does not imply that they are forbidden. More precisely,
they may be forbidden by some additional physical principle taken together with the principle of relativity, but they 
are {\em not forbidden by the principle of relativity alone}.

Let us also demonstrate that the results above reduce to familiar physics
when the potential $U$ is a positive constant. In this case, the force on the
right-hand side of (\ref{nikolic:book13}) vanishes. 
It is convenient to fix the constant $m$ in (\ref{fix}) such that 
\begin{equation}\label{U=m/2}
 U=\frac{m}{2} ,
\end{equation}
because then (\ref{nikolic:book12}) and (\ref{nikolic:book4})-(\ref{nikolic:book5})
imply that $s$ is equal to the proper time $\tau$.
To fully understand the physical meaning of (\ref{U=m/2}), it is also useful
to work out the nonrelativistic limit of (\ref{nikolic:book11}). In this limit
$\tau \simeq t$ and
\begin{equation}
 -\frac{m}{2} \dot{X}^{\mu} \dot{X}_{\mu} \simeq -\frac{m}{2} +
\frac{m\dot{\bf X}^2}{2} ,
\end{equation}
so (\ref{nikolic:book11}) reduces to
\begin{equation}\label{nikolic:book11nr}
 A \simeq \int dt \left[\frac{m\dot{\bf X}^2}{2} - m \right] .
\end{equation}
The first term is the usual action for a free nonrelativistic particle with mass $m$,
while the second term is a constant which does not affect the equations of motion.
The second term shows that the potential energy of particle at rest is $m$ 
(not $m/2$ as one might naively think from (\ref{U=m/2})).

\subsection{Canonical momentum and the Hamilton-Jacobi formulation}
 
Physics defined by (\ref{nikolic:book11})-(\ref{nikolic:book12}) can also be described by introducing
the canonical momentum
\begin{equation}\label{nikolic:book14}
 P_{\mu}=\frac{\partial L}{\partial \dot{X}^{\mu}} ,
\end{equation}
where  
\begin{equation}\label{nikolic:book15}
 L(X,\dot{X})=- \frac{m}{2} \dot{X}^{\mu} \dot{X}_{\mu} - U(X)   .
\end{equation}
This leads to
\begin{equation}\label{nikolic:book16}
 P^{\mu}=-m\dot{X}^{\mu} .
\end{equation}
The canonical Hamiltonian is
\begin{equation}\label{nikolic:book17}
H(P,X)=P_{\mu}\dot{X}^{\mu}-L=-\frac{P^{\mu}P_{\mu}}{2m}+U(X) .
\end{equation}
Note that this Hamiltonian is {\em not} the energy of the particle. In particular,
while particle energy transforms as a time-component of a spacetime vector,
the Hamiltonian above transforms as a scalar. This is a consequence of the fact 
$\dot{X}^{\mu}$ is not a derivative with respect to time $x^0$, but a derivative
with respect to the scalar $s$.

The constraint (\ref{nikolic:book12}) now can be written as
\begin{equation}\label{nikolic:book18}
 P^{\mu}P_{\mu}=2mU(X) .
\end{equation}
In relativity, it is customary to {\em define} the invariant mass $M$ through
the identity $P^{\mu}P_{\mu}\equiv M^2$. This shows that the mass depends
on $X$ as
\begin{equation}\label{nikolic:book19}
 M^2(X)=2mU(X) .
\end{equation}
Since $U(X)$ may change sign as $X$ varies, we see that the particle may, e.g., 
start motion as an ``ordinary'' massive particle ($M^2>0$) and later evolve into a
tachyon ($M^2<0$). The usual proof that an ``ordinary'' particle cannot reach
(or exceed) the velocity of light involves an assumption that the mass is a constant.
When mass is not a constant, or more precisely when $M^2$ can change sign,
then particle can reach and exceed the velocity of light.
 
The existence of the Hamiltonian allows us to formulate classical relativistic mechanics
with the relativistic Hamilton-Jacobi formalism.
One introduces the scalar Hamilton-Jacobi function $S(x,s)$ satisfying the 
Hamilton-Jacobi equation
\begin{equation}\label{nikolic:book21}
H(\partial S,x)=-\frac{\partial S}{\partial s} .
\end{equation}
Comparing (\ref{nikolic:book18}) with (\ref{nikolic:book17}), we see that the constraint
(\ref{nikolic:book18}) can be written as
 \begin{equation}\label{nikolic:book20}
  H(P,X)=0 .
 \end{equation}
The constraint (\ref{nikolic:book20}) implies that the right-hand side of (\ref{nikolic:book21})
must vanish, i.e., that $S(x,s)=S(x)$.
Hence (\ref{nikolic:book21}) reduces to $H(\partial S,x)=0$, which
in an explicit form reads
\begin{equation}\label{nikolic:book22}
 -\frac{(\partial^{\mu}S) (\partial_{\mu}S)}{2m}+U(x) =0 .
\end{equation}
The solution $S(x)$ determines the particle momentum 
\begin{equation}\label{nikolic:book23}
 P^{\mu}=\partial^{\mu}S(X) ,
\end{equation}
which, through (\ref{nikolic:book16}), determines the particle trajectory
\begin{equation}\label{nikolic:book24}
 \frac{dX^{\mu}(s)}{ds}=-\frac{\partial^{\mu}S(X(s))}{m} .
\end{equation}

\subsection{Generalization to many particles}
\label{SEC4.3}

Now, let us briefly generalize all this to the case of many particles.
We study the dynamics of $n$ trajectories $X_a^{\mu}(s)$, $a=1,\ldots,n$, 
parameterized by a single parameter $s$.
In the general action (\ref{nikolic:book7}), the velocity-dependent terms
generalize as follows
\begin{equation}
\dot{X}^{\mu}C_{\mu} \rightarrow \sum_{a=1}^n \dot{X}_a^{\mu} C_{a\mu} ,
\end{equation}
\begin{equation}
\dot{X}^{\mu}\dot{X}^{\nu}C_{\mu\nu} \rightarrow 
\sum_{a=1}^n  \sum_{b=1}^n \dot{X}_a^{\mu} \dot{X}_b^{\nu}C_{ab\mu\nu} .
\end{equation}
Since the scalar potential is our main concern, we consider trivial vector and tensor potentials
$C_{a\mu}=0$ and $C_{ab\mu\nu}=c_a\delta_{ab}\eta_{\mu\nu}$, respectively, where
$c_a$ are constants.
Thus, Eqs.~(\ref{nikolic:book11})-(\ref{nikolic:book12}) generalize to
\begin{equation}\label{nikolic:book25}
 A=-\int ds \left[  \sum_{a=1}^n \frac{m_a}{2}
\dot{X}_a^{\mu} \dot{X}_{a\mu} + 
U(X_1,\ldots,X_n) \right]  ,
\end{equation}
\begin{equation}\label{nikolic:book26}
 \sum_{a=1}^n m_a \dot{X}_a^{\mu}\dot{X}_{a\mu}=2U(X_1,\ldots,X_n) ,
\end{equation}
where $m_a=mc_a$ and $c_a$ are dimensionless.
The relativistic Newton equation (\ref{nikolic:book13}) generalizes to
\begin{equation}\label{nikolic:book27}
 m_a\frac{d^2X_a^{\mu}(s)}{ds^2}=\partial_a^{\mu}U(X_1(s),\ldots,X_n(s)) .
\end{equation}
In general, from (\ref{nikolic:book27}) we see that that the force on the particle $a$
at the spacetime position $X_a(s)$ depends on positions of all other particles
for the same $s$. In other words, the forces on particles are nonlocal. Nevertheless,
since $s$ is a scalar, such nonlocal forces are compatible with the principle of relativity;
the nonlocal equation of motion (\ref{nikolic:book27}) is relativistic covariant.
Thus we see that {\em relativity and nonlocality are compatible with each other}.
Even though for each $s$ there may exist a particular ($s$-dependent) 
Lorentz frame with respect to which
the force between two particles is instantaneous, such a Lorentz frame is by no means
special or ``preferred''. Instead, such a particular Lorentz frame is determined by covariant
equations of motion supplemented by a particular choice of initial
conditions $X_a^{\mu}(0)$.

Note also that the phenomena of {\em nonlocal forces
between particles} and {\em particle motions faster than light} are independent of each other.
The force (\ref{nikolic:book27}) becomes local when
\begin{equation}\label{nikolic:book28}
 U(X_1,\ldots,X_n)=U_1(X_1)+ \cdots + U_n(X_n) ,
\end{equation}
in which case (\ref{nikolic:book27}) reduces to
\begin{equation}\label{nikolic:book29}
 m_a\frac{d^2X_a^{\mu}(s)}{ds^2}=\partial_a^{\mu}U_a(X_a(s)) .
\end{equation}
Thus we see that particle motions faster than light ($U_a<0$)
are possible even when the forces are local.
Similarly, $U(X_1,\ldots,X_n)$ may be such that particles move only slower than light,
but that the forces are still nonlocal. 

The Hamilton-Jacobi formalism can also be generalized to the many-particle case.
In particular, Eqs.~(\ref{nikolic:book22}) and (\ref{nikolic:book24}) generalize to
\begin{equation}\label{nikolic:book30}
 -\sum_{a=1}^n \frac{(\partial_a^{\mu}S) (\partial_{a\mu}S)}{2m_a}
+U(x_1,\ldots, x_n) =0 ,
\end{equation}
\begin{equation}\label{nikolic:book31}
 \frac{dX_a^{\mu}(s)}{ds}=-\frac{\partial_a^{\mu}S(X_1(s),\ldots, X_n(s))}{m_a} ,
\end{equation}
respectively.
In the local case (\ref{nikolic:book28}), the solution of (\ref{nikolic:book30}) can be written in the form
\begin{equation}\label{nikolic:book32}
 S(x_1,\ldots x_n)=S_1(x_1)+ \cdots + S_n(x_n) ,
\end{equation}
so (\ref{nikolic:book31}) reduces to
\begin{equation}\label{nikolic:book33}
 \frac{dX_a^{\mu}(s)}{ds}=-\frac{\partial_a^{\mu}S_a(X_a(s))}{m_a} .
\end{equation}

\subsection{Analogy with Newtonian mechanics and the notion of time}

It is evident that relativistic mechanics with a scalar potential in spacetime is formally
analogous to Newtonian mechanics with a scalar potential in space.
Essentially, the transition from Newtonian mechanics to relativistic mechanics
is accomplished by the following correspondence:
\begin{eqnarray}
& {\bf x}=(x^1,x^2,x^3) \rightarrow x=(x^0,x^1,x^2,x^3) , &
\nonumber \\
& t \rightarrow s . &
\end{eqnarray}
In this sense, (\ref{Newton}) is analogous to (\ref{nikolic:book27}), 
(\ref{Ham}) is analogous  to (the obvious many-particle variant of) (\ref{nikolic:book17}),
(\ref{momenta}) is analogous to (the obvious many-particle variant of) (\ref{nikolic:book16}), 
(\ref{HamJac}) is analogous to (the obvious many-particle variant of) (\ref{nikolic:book21}), 
(\ref{traj}) is analogous to (\ref{nikolic:book31}),
and (\ref{Hamconstr}) is analogous to (the obvious many-particle variant of) (\ref{nikolic:book20}).
The relativistic analogue of (\ref{dt2}) is provided by Eq.~(\ref{ds2}) in the Appendix.

Also, even though $s$ is an auxiliary parameter that only serves to parameterize the
trajectories in spacetime, it can be measured indirectly in the same sense as $t$ can be
measured indirectly in Newtonian mechanics, as explained in Sec.~\ref{SEC2}. 
Namely, if at least one of the $4n$ functions $X^{\mu}_a(s)$ is periodic in $s$,
then the number of periods can be interpreted as a measure of
elapsed $s$.
Owing to this analogy between absolute Newton time $t$ and the parameter $s$,
it is justified to think of $s$ as an absolute time in relativistic mechanics.

Thus we see that in relativistic physics one can introduce two different notions of
invariant time. One is the proper time measured by a ``clock'' periodic in the proper time $\tau$,
while the other is the absolute time measured by a ``clock'' periodic in $s$. 
These two invariant times coincide
in many cases of practical interest, but in general they are different.
For a more formal relation between these two notions of time see also
the Appendix.

\section{Example: Relativistic Bohmian mechanics}
\label{SEC5}

\subsection{Klein-Gordon equation of a single particle}

Consider a particle which is free on the classical level,
i.e., a particle classically described by the action (\ref{nikolic:book11})
with a constant scalar potential (\ref{U=m/2}).
The constraint (\ref{nikolic:book18}) becomes
\begin{equation}\label{nikolic:consc}
 P^{\mu}P_{\mu}-m^2 =0 ,
\end{equation}
implying that $m$ is the mass of the particle.

In quantum mechanics, the momentum $P_{\mu}$ becomes the operator $\hat{P}_{\mu}$
satisfying the canonical commutation relations
\begin{equation}\label{nikolic:e3}
[x^{\mu},\hat{P}_{\nu}]=-i\eta^{\mu}_{\nu} ,
\end{equation}
where we work in units $\hbar=1$. These commutation relations are satisfied
by taking
\begin{equation}\label{nikolic:e2}
\hat{P}_{\nu}=i\partial_{\nu} .
\end{equation}

The quantum analog of the classical constraint (\ref{nikolic:consc}) is
\begin{equation}\label{nikolic:consq}
 [\hat{P}^{\mu}\hat{P}_{\mu}-m^2]\psi(x) =0 ,
\end{equation}
where $\psi(x)$ is the wave function.
Eq.~(\ref{nikolic:consq}) is nothing but the Klein-Gordon equation
\begin{equation}\label{nikolic:KG}
 [\partial^{\mu}\partial_{\mu}+m^2]\psi(x) =0 .
\end{equation}
By writing $\psi=Re^{iS}$, where $R$ and $S$ are real functions,
the complex Klein-Gordon equation (\ref{nikolic:KG}) is equivalent to a set of
two real equations
\begin{equation}\label{nikolic:cont}
\partial^{\mu}(R^2\partial_{\mu}S)=0,
\end{equation}
\begin{equation}\label{nikolic:HJ}
-\frac{(\partial^{\mu}S)(\partial_{\mu}S)}{2m} +\frac{m}{2} +Q=0,
\end{equation}
where (\ref{nikolic:cont}) is the conservation equation and
\begin{equation}\label{nikolic:Q=}
Q=\frac{1}{2m}\frac{\partial^{\mu}\partial_{\mu}R}{R} .
\end{equation}

\subsection{Klein-Gordon equation of many particles}

Now let us generalize it to the case of $n$ identical particles without spin,
with equal masses $m_a=m$.
The wave function $\psi$ satisfies $n$ Klein-Gordon equations
\begin{equation}\label{nikolic:KGn}
(\partial_a^{\mu}\partial_{a\mu}+m^2)\psi(x_1,\ldots ,x_n)=0 ,
\end{equation}
one for each $x_a$.
Equation (\ref{nikolic:KGn}) implies 
\begin{equation}\label{nikolic:KGs}
\left( \sum_a\partial_a^{\mu}\partial_{a\mu}+nm^2 \right)
\psi(x_1,\ldots ,x_n)=0 .
\end{equation}
Next we write $\psi=Re^{iS}$, where $R$ and $S$ are real
functions. Equation (\ref{nikolic:KGs})
is then equivalent to a set of two real equations
\begin{equation}\label{nikolic:contn}
\sum_a\partial_a^{\mu}(R^2\partial_{a\mu}S)=0,
\end{equation}
\begin{equation}\label{nikolic:HJn}
-\frac{\sum_a(\partial_a^{\mu}S)(\partial_{a\mu}S)}{2m} +\frac{nm}{2} +Q=0,
\end{equation}
where
\begin{equation}\label{nikolic:Q}
Q=\frac{1}{2m}\frac{\sum_a\partial_a^{\mu}\partial_{a\mu}R}{R} .
\end{equation}

\subsection{The Bohmian interpretation}

The crucial observation is that the quantum equation (\ref{nikolic:HJ}) has the same form
as the classical equation (\ref{nikolic:book22}), provided that we make the identification
\begin{equation}\label{nikolic:u-q}
 U(x) = \frac{m}{2}+Q(x) .
\end{equation}
The first term on the right-hand side of (\ref{nikolic:u-q})
is the classical potential (\ref{U=m/2}), while the second
term is the quantum potential. (Recall that we work in units $\hbar=1$. 
In units in which $\hbar\neq 1$, it is easy to show that (\ref{nikolic:Q=}) attains an additional factor
$\hbar^2$, showing that the quantum potential $Q$ vanishes in the classical limit.)
This suggests the Bohmian interpretation, according to which (\ref{nikolic:HJ}) is the quantum
Hamilton-Jacobi equation and the particle has the trajectory given by (\ref{nikolic:book24})
\begin{equation}\label{nikolic:book24q}
 \frac{dX^{\mu}(s)}{ds}=-\frac{\partial^{\mu}S(X(s))}{m} .
\end{equation}
From (\ref{nikolic:book24q}), (\ref{nikolic:HJ}), and the identity
\begin{equation}
\frac{d}{ds}=\frac{dX^{\mu}}{ds}\partial_{\mu} ,
\end{equation}
one finds a quantum variant of (\ref{nikolic:book13})
\begin{equation}\label{nikolic:book13q}
 m\frac{d^2X^{\mu}(s)}{ds^2}=\partial^{\mu}Q(X(s)) .
\end{equation}
It is well-known that Bohmian trajectories satisfying (\ref{nikolic:book24q}) may lead 
to superluminal velocities \cite{BMbook1,BMbook2}. As discussed in more detail
in \cite{niktorino}, such superluminal velocities do not lead to inconsistencies
or conflicts with observations.

Now the generalization to $n$ particles is straightforward. Essentially, 
all equations above are rewritten such that each quantity having the index $\mu$ receives
an additional index $a$. In particular, Eqs.~(\ref{nikolic:book24q}) are (\ref{nikolic:book13q}) 
generalize to
\begin{equation}\label{nikolic:book24qn}
 \frac{dX_a^{\mu}(s)}{ds}=-\frac{\partial_a^{\mu}S(X_1(s),\ldots,X_n(s))}{m} ,
\end{equation}
\begin{equation}\label{nikolic:book13qn}
 m\frac{d^2X_a^{\mu}(s)}{ds^2}=\partial_a^{\mu}Q(X_1(s),\ldots,X_n(s)) ,
\end{equation}
respectively.
In general, particles have nonlocal influences on each other, in exactly the same way
as in classical relativistic mechanics as studied in Sec.~\ref{SEC4.3}.

The compatibility of these particle trajectories with the probabilistic predictions
of quantum theory is studied elsewhere \cite{nikbosfer,nikijqi,nikqft,niktorino}. 

\section{Conclusion}
\label{SEC6}

In this paper, we have systematically developed a relativistic-covariant formalism
describing kinematics and dynamics of classical particles in a background scalar potential. 
The scalar potential promotes the particle mass to a dynamical quantity. 
The principle of relativity does not restrict the sign of the potential, which provides
a natural way to describe particles with superluminal velocities.
In the many-particle case, the formalism also allows nonlocal relativistic 
interactions. In many respects, the formalism is analogous to the formalism
of nonrelativistic Newtonian mechanics. In particular, the scalar parameter
that parameterizes relativistic particle trajectories is analogous to the Newton
absolute time.

As an example, we have also demonstrated that relativistic Bohmian mechanics
of spin-0 particles can be viewed as a special case of the general formalism
for classical relativistic particles. This provides a 
deeper view of relativistic Bohmian mechanics and makes the understanding 
of the physical meaning of it more complete.

\section*{Acknowledgements}

This work was supported by the Ministry of Science of the
Republic of Croatia under Contract No.~098-0982930-2864.

\appendix

\section{Generalized proper time}
\label{APP}

In this section we study the relation between the parameter $s$ and
the proper time $\tau$ defined 
for timelike trajectories as
\begin{equation}
 d\tau^2=\eta_{\mu\nu}dX^{\mu}dX^{\nu} . 
\end{equation}
Using (\ref{nikolic:book19}), we see that the constraint (\ref{nikolic:book12})
can be written as
\begin{equation}\label{A1}
 \eta_{\mu\nu}\frac{dX^{\mu}}{ds}\frac{dX^{\nu}}{ds} = \frac{M^2(X)}{m^2} .
\end{equation}
Introducing the conformal metric
\begin{equation}
 \tilde{\eta}_{\mu\nu}(x) \equiv \frac{m^2}{M^2(x)}\eta_{\mu\nu},
\end{equation}
(\ref{A1}) can be written as
\begin{equation}\label{A3}
 \tilde{\eta}_{\mu\nu}(X) \frac{dX^{\mu}}{ds}\frac{dX^{\nu}}{ds} = 1 .
\end{equation}
Thus, we see that the parameter $s$ can be interpreted as a generalized proper
time defined with respect to the conformal metric as
\begin{equation}
 ds^2=\tilde{\eta}_{\mu\nu}(X)dX^{\mu}dX^{\nu} .
\end{equation}
In particular, when the mass $M$ is positive and constant, then it is convenient to fix the positive constant 
$m$ in (\ref{fix}) such that $M=m$, in which case $\tilde{\eta}_{\mu\nu}(X)=\eta_{\mu\nu}$
and $ds^2=d{\tau}^2$. The generalized proper time is appropriate for a generalization of
the notion of proper time to the case in which mass is a dynamical quantity. In particular,
unlike the ordinary proper time, the generalized proper time is well-defined even when
the particle trajectory is spacelike or lightlike. The quantity $ds^2$ 
is always positive along the trajectory, even when the particle trajectory is spacelike or lightlike.

Formally, this can also be generalized to the case of many particles. 
We introduce the condensed notation $A\equiv(a,\mu)$, so that the coordinates $X^{\mu}_a$ can 
be thought of as coordinates $X^A$ in the $4n$-dimensional configuration space
with the metric
\begin{equation}
 \eta_{AB}=\delta_{ab}\eta_{\mu\nu} .
\end{equation}
Introducing the conformal metric
\begin{equation}
\tilde{\eta}_{AB}(X)=\frac{m_a}{2U(X)} \delta_{ab}\eta_{\mu\nu},  
\end{equation}
where $X\equiv \{ X^A \}$,
the constraint (\ref{nikolic:book26}) can be written as
\begin{equation}\label{nikolic:book26-A}
\tilde{\eta}_{AB}(X) \frac{dX^A}{ds}\frac{dX^B}{ds}=1 ,
\end{equation}
where the summation over repeated indices is understood.
Thus we see that $s$ can be thought of as a generalized proper time defined as
\begin{equation}\label{ds2}
 ds^2=\tilde{\eta}_{AB}(X) dX^A dX^B .
\end{equation}
This is analogous to the nonrelativistic relation (\ref{dt2}).

As a special case, consider $n$ free particles, each with a constant positive mass $m_a$.
In this case, (\ref{U=m/2}) generalizes to
\begin{equation}
 U=\sum_{a=1}^{n} \frac{m_a}{2}. 
\end{equation}
Eq.~(\ref{nikolic:book26}) becomes
\begin{equation}\label{nikolic:book26const}
 \sum_{a=1}^n m_a \frac{dX_a^{\mu}}{ds} \frac{dX_{a\mu}}{ds}= \sum_{a=1}^n m_a .
\end{equation}
Since $dX_a^{\mu}dX_{a\mu}=d\tau_a^2$, (\ref{nikolic:book26const})
implies that
\begin{equation}\label{APPaverage}
 ds^2=\sum_{a=1}^{n}w_a d\tau_a^2 ,
\end{equation}
where
\begin{equation}\label{APPw}
w_a = \frac{m_a}{\sum_{b=1}^{n} m_{b}} .
\end{equation}
Eqs.~(\ref{APPaverage})-(\ref{APPw}) 
show that $ds^2$ is the {\em average} quadratic differential proper time,
where average is defined with weight factors $w_a$ satisfying $\sum_{a=1}^{n}w_a =1$.

\end{document}